# A Trichotomy for Regular Simple Path Queries on Graphs

Guillaume Bagan
INRIA, France
guillaume.bagan@inria.fr

Angela Bonifati
Lille 1 University & INRIA, France
angela.bonifati@inria.fr

Benoit Groz
Tel-Aviv University, Israel
benoit.groz@inria.fr

## ABSTRACT

Regular path queries (RPQs) select nodes connected by some path in a graph. The edge labels of such a path have to form a word that matches a given regular expression. We investigate the evaluation of RPQs with an additional constraint that prevents multiple traversals of the same nodes. Those regular simple path queries (RSPQs) find several applications in practice, yet they quickly become intractable, even for basic languages such as $(aa)^*$ or $a^*ba^*$.

In this paper, we establish a comprehensive classification of regular languages with respect to the complexity of the corresponding regular simple path query problem. More precisely, we identify the fragment that is maximal in the following sense: regular simple path queries can be evaluated in polynomial time for every regular language $L$ that belongs to this fragment and evaluation is NP-complete for languages outside this fragment. We thus fully characterize the frontier between tractability and intractability for RSPQs, and we refine our results to show the following trichotomy: Evaluations of RSPQs is either $AC^0$, NL-complete or NP-complete in data complexity, depending on the regular language $L$. The fragment identified also admits a simple characterization in terms of regular expressions.

Finally, we also discuss the complexity of the following decision problem: decide, given a language $L$, whether finding a regular simple path for $L$ is tractable. We consider several alternative representations of $L$: DFAs, NFAs or regular expressions, and prove that this problem is NL-complete for the first representation and PSPACE-complete for the other two. As a conclusion we extend our results from edge-labeled graphs to vertex-labeled graphs and vertex-edge labeled graphs.

## 1. INTRODUCTION

The reachability problem for graphs (finding a path between two nodes) represents one of the oldest problems in computer science for which very efficient algorithms have been conceived. However, for many real-world problems, constraints on the path need to be considered and, as a consequence, the reachability problem can become computationally hard. Problems on regular paths are among the most studied class of constrained path problems. In these problems the edge labels along the path must form a word belonging to a given regular language. For graph databases, such problems have been considered in the context of regular path queries (RPQs). Given a language $L$ and two vertices in a database graph, a regular path query selects pairs of nodes connected by a path whose edge labels form a word in $L$. Graph databases and RPQs have been investigated starting from the late 80s [1, 5, 8, 9, 10, 13, 14, 18, 27, 29], and are now again in vogue due to their wide application scenarios, e.g. in social networks [35], biological and scientific databases [26, 32], and the Semantic Web [17]. Regular path queries allow to traverse the same nodes multiple times, whereas regular simple path queries (RSPQs) permit to traverse each vertex only once. From a theoretical viewpoint, the former notion has overridden the latter, mainly for complexity reasons. Indeed, RPQs are computable in time polynomial in both query and data complexity (combined complexity), while the evaluation of RSPQs is NP-complete even for fixed basic languages such as $(aa)^*$ or $a^*ba^*$ [29]. RSPQs, however, are desired in many application scenarios [26, 32, 6, 24, 22, 39], such as transportation problems, VLSI design, metabolic networks, DNA matching and routing in wireless networks. As a further example, the problem of finding subgraphs matching a graph pattern can be generalized to use regular expressions on pattern edges [14]. Such queries may also enforce the condition that their matched vertices are distinct. Additionally, regular simple paths have been recently considered in SPARQL 1.1 queries exhibiting property paths. In particular, recent studies on the complexity of property paths in SPARQL [3, 28] have highlighted the hardness of the semantics proposed by W3C to evaluate such paths in RDF graphs. Roughly speaking, according to the semantics considered in [28], the evaluation of expressions under Kleene-star closure imposes that the involved path is simple, whereas the evaluation of the remaining expressions allows to traverse the same node multiple times. As such, the semantics studied in [28] is an hybrid between regular paths and regular simple paths semantics.

*Contributions.* In this paper, we address the long standing open question [29, 6] of characterizing exactly the maximal class of regular languages for which RSPQs are tractable. By "tractable" we mean computable in time polynomial in the size of the database. Precisely, we establish a comprehensive classification of the complexity of RSPQs for a fixed regular language $L$: given a edge-labeled graph $G$ and two vertices $x$ and $y$, is there a simple path from $x$ to $y$ whose edge labels form a word of $L$? A first step towards this important issue has been made in [29]. They exhibit a tractable fragment: the class of languages closed by subword. However, their fragment is not maximal.

Our contributions can be detailed as follows. We introduce a class of languages, named $trC$, for which RSPQs are computable in polynomial time, and even in NL. We then show that this fragment is maximal as the RSPQ problem is NP-complete for every regular language that does not belong to $trC$. Consequently, we characterize, under the hypothesis NL $\neq$ NP that is actually weaker than PTIME $\neq$ NP, the frontier between tractability and intractability for this problem. Additionally, $trC$ also represents the maximal class for which finding a shortest path that satisfies a RSPQ is tractable. We note that we focus on data complexity as we assume that the language $L$ is fixed. At this point, the chart of the classification of the languages is not yet complete. Therefore, we refine our results to show the following trichotomy: the RSPQ problem is either $AC^0$, NL-complete or NP-complete.

We discuss the complexity to decide, given a language $L$, whether the RSPQ problem for $L$ is tractable. We consider several alternative representations of $L$: DFAs, NFAs or regular expressions. We prove that this problem is NL-complete for the first representation and PSPACE-complete for the two others.

Next, we give a characterization of the tractable fragment $trC$ for edge-labeled graphs in term of regular expressions. Moreover, $trC$ is closed by union and intersection and languages in $trC$ are aperiodic i.e. can be expressed by first-order formulas [37].

The above results hold for the common definition of database graphs, i.e. edge-labeled graphs. However, it seems natural to take into consideration both queries on top of vertices labels and queries on top of vertices and edges labels. As an example, a Google Maps user may be interested to specify as a condition a regular expression that enforces a stop over in a given city and avoids another city while preferring certain types of roads. For such a reason, we focus on two other models: vertex-labeled graphs or vertex-edge-labeled graphs (where both vertices and edges are labeled). Surprisingly, for some languages, the RSPQ problem is simpler on vertex-labeled graphs than on edge-labeled graphs. With $L = (ab)^*$ for instance, RSPQ is polynomial for vertex-labeled graphs and NP-complete for edge-labeled graphs. Vertex-edge-labeled graphs obviously generalize both edge-labeled graphs and vertex-labeled graphs. Furthermore, we can adapt our results to prove, for these two models, a classification of the same kind as the one shown for vertex-labeled graphs: the RSPQ problem is either $AC^0$, NL-complete or NP-complete.

As a final contribution, we have obtained two minor results. First, we have attempted to study the parametrized complexity of tractable RSPQs queries when the parameter is the size of the query. However, we obtained a partial result: we prove that the problem is FPT for the class of finite languages. Moreover, we prove that the problem is also FPT for the class of all regular languages when the parameter is the size of the path. As a second result, we prove that the problem RSPQis polynomial w.r.t. combined complexity on graphs of bounded directed treewidth. This is actually a straightforward generalization of a result of [23].

*Related Work.* Regular path queries express ways to evaluate regular expression patterns on database graph models [1, 5, 8, 9, 10, 13, 14, 18, 27, 29] or tree-structured models, such as XML [11]. While the regular path problem has been extensively studied in the literature, the regular simple path problem has received less attention in both the database and graph communities. Besides the works on regular paths, there have been studies on finding paths with some constraints. In particular, Lapaugh et al. [25] prove that finding simple paths of even length is polynomial for non directed graphs and NP-complete for directed graphs. This study has been extended in [4] by considering paths of length $i \mod k$. Similarly, finding $k$ disjoint paths with extremities given as input is polynomial for non directed graphs [34] and NP-complete for directed graphs [16]. Mendelzon and Wood [29] show that the regular simple path problem is NP-complete in the general case. However, they show that the problem can be decided in polynomial time for subword-closed languages. They also show that the problem becomes polynomial under some restrictions on the size of cycles of both graph and automaton. A subsequent paper [31] proves the polynomiality for the class of outerplanar graphs. Barrett et al. [6] extend this result, proving that the regular simple path problem is polynomial w.r.t. combined complexity for graphs of bounded treewidth. Let us also observe that the existence of a regular simple path between two vertices is MSO-definable, and therefore a well-known result of Courcelle [12] already implies the same result but w.r.t. data complexity only. Barrett et al. [6] also show that the problem is NP-complete for the class of grid graphs even when the language is fixed. Practical algorithms for regular simple paths on large graphs have been proposed in [22, 24].

Regular simple paths have been also investigated in the context of SPARQL property paths with the semantics proposed in a working draft of SPARQL 1.1. Notice that such semantics of SPARQL property paths doesn't exactly correspond to regular simple paths queries. Losemann and Martens [28] and Arenas et al. [3] investigate the complexity of evaluating such property paths. They show that the evaluation is NP-complete in several cases, along with exhibiting cases in which it is polynomial. More precisely, Losemann and Martens consider different fragments of regular expressions and classify them with respect to the complexity of evaluating SPARQL property paths. Both papers also show that counting the number of paths that match a regular expression (which is permitted by the working draft) is hard in many cases.

## 2. PRELIMINARIES

For the rest of the paper, $\Sigma$ always denotes a finite alphabet. We use the notation $[n]$ to denote the set of integers $\{1, \ldots, n\}$. Given a word $w$ and a language $L$, $w^{-1}L = \{w' : ww' \in L\}$.

*Complexity:* $\mathrm{NL}, \mathrm{P}, \mathrm{NP}, \mathrm{PSPACE}$ refer to the classical classes of complexity [33]. The reductions we consider are many-to-one logspace reductions [33] and completeness of problems are under this type of reductions. The class $\mathrm{AC}^0$ refers to uniform $\mathrm{AC}^0$ that is equivalent to $\mathrm{FO}(+, \times)$ or $\mathrm{FO}(\mathrm{BIT}, <)$ [21]. For definition of FPT, see [15].

*Graphs:* In our paper, we essentially consider db-graphs even if we consider vertex-labeled graphs and evl-graphs (graphs where both vertices and edges are labeled). A db-graph is a tuple $G = (V, \Sigma, E)$ where $V$ is a set of vertices, $\Sigma$ is a set of labels and $E \subseteq V \times \Sigma \times V$ is a set of edges labeled by symbols of $\Sigma$. A path $p$ of a db-graph $G$ from $x$ to $y$ is a sequence $(v_1 = x, a_1, \ldots, v_k, a_k, v_{k+1} = y)$ such for each $i \in [k+1]$, $v_i$ is a vertex in $G$ and for each $i \in [k]$, $(v_i, a_i, v_{i+1})$ is an edge in $G$. A path $p$ is simple if all vertices $v_i$ in $p$ are distinct. Given a language $L \subseteq \Sigma^*$, $p$ is an $L$-labeled path if $a_1 \ldots a_k \in L$.

*Automata:* Let $L$ be a regular language. We denote by $A_L = (Q_L, i_L, F_L, \Delta_L)$ the minimal DFA for $L$, and by $M$ the number of states $M = |Q_L|$ in $A_L$. We assume that $A_L$ is complete i.e. $\Delta_L$ is a total function, so that in general $A_L$ may have a sink state. For any $q \in Q, w \in \Sigma^*$, $\Delta_L(q, w)$ denotes the state obtained when reading $w$ from $q$. Finally, $\mathscr{L}_q$ denotes the set of all words accepted from $q$. For every state $q$ we denote by $Loop(q)$ the set of all non empty words that allow to loop on $q$: $Loop(q) = \{w \in \Sigma^+ \mid \Delta(q, w) = q\}$. Strongly connected components of (the graph of) $A_L$ are simply called components. There are many definitions of aperiodic languages [37]. We use the following. A language $L$ is aperiodic if it is regular and its minimal automaton $A_L$ satisfies the following property for every state $q \in Q_L$, integer $k \geq 1$ and word $w \in \Sigma^*$: $\Delta_L(q, w^k) = q \Rightarrow \Delta(q, w) = q$. As consequence, for every state $q \in Q_L$ and word $w \in \Sigma^*$, $\Delta_L(q, w^{M+1}) = \Delta_L(q, w^M)$.

*RSPQ:* Given a class $\mathcal{L}$ of regular languages and a class $\mathcal{G}$ of db-graphs, we define the following problem:

---
**RSPQ**($\mathcal{L}, \mathcal{G}$)

**Input:** a language $L \in \mathcal{L}$,
a db-graph $G = (V, \Sigma, E) \in \mathcal{G}$,
and two vertices $x, y \in V$
**Question:** is there a simple $L$-labeled path from $x$ to $y$?

---

The encoding of the language $L$ will be specified when required. We denote by "All" the class of db-graphs, RSPQ($\mathcal{L}$) means RSPQ($\mathcal{L}$, All). For a fixed language $L$, we use RSPQ($L, \mathcal{G}$) to denote RSPQ($\{L\}, \mathcal{G}$). Since $L$ is fixed, we focus on data complexity. Notice that the representation of $L$ does not matter here. Although we consider the boolean version of the problem, namely deciding the existence of a path, our algorithms actually also return a (shortest) simple $L$-labeled path.

Given a regular language $L$, our main question is to give a criterion to decide whether RSPQ($L$) is tractable (i.e. decidable in polynomial time) or not (i.e NP-complete). We address this question in the next and following sections.

EXAMPLE 1. *As an introductory example, consider the language $L = a^*(bb^+ + \varepsilon)c^*$. We wish to decide whether there exists a simple path from $x$ to $y$ labeled by $L$, given two vertices $x, y$ of a db-graph $G$. It is not absolutely trivial that this problem can be solved efficiently: the problem has indeed been proved NP-complete for the language $a^*bc^*$. Yet we can give a polynomial algorithm for $L$.*

*First, we check whether $y$ can be reached from $x$ by a (non-necessarily simple) path labeled in $a^*c^*$. In that case we can obtain a simple such path since the path obtained after eliminating the loops is still labeled in $a^*c^*$.*

*Assume now that no $a^*c^*$-labeled path can connect $x$ to $y$. Then we can similarly check if there exists a simple $a^*b^kc^*$-labeled path from $x$ to $y$ for $k \in \{2, 3\}$: we check one after another each possible assignment for the $k$ middle $b$-labeled edges: if the initial $b$-labeled edge can be reached from $x$ via an $a^*$-labeled path (avoiding the summits of the $b$-edges) and if the final $b$-labeled edge can reach $y$ through a $c^*$-labeled path, then we obtain a simple $a^*b^kc^*$-labeled path from $x$ to $y$. This is because we assumed there is no $a^*c^*$-labeled path from $x$ to $y$, so the $a$- and $c$-labeled edges cannot intersect. Observe that the number of possible assignment for $k$ edges ($k \leq 3$) is polynomial.*

*Let us now assume w.l.o.g. that there is no $a^*b^kc^*$-*

labeled path from $x$ to $y$ for $k \in \{0, 2, 3\}$. We can show that there exists a simple $L$-labeled path from $x$ to $y$ if and only if there exist six nodes $v_1, v_2, v_3, v_4, v_5, v_6$, all distinct except that $v_3$ may equal $v_4$, and for which we can find simultaneously:

- a $b$-labeled edge from $v_1$ to $v_2$, from $v_2$ to $v_3$, from $v_4$ to $v_5$, and from $v_5$ to $v_6$.
- an $a^*$-labeled path from $x$ to $v_1$ avoiding all other $v_i$ $(i > 1)$
- a $b^*$-labeled path from $v_3$ to $v_4$ of which all nodes (but the first and last) avoid $S_a$.
- a $c^*$-labeled path from $v_6$ to $y$ of which all nodes (but the first) avoid $S_a$ and $S_b$.

where the set $S_a$ contains exactly all $v_i$ plus all positions reachable from $x$ by some $a^*$-labeled path avoiding all $v_i$, and the set $S_b$ contains exactly all $v_i$ plus all positions reachable from $v_3$ through a $b^*$-labeled path that avoids all nodes of $S_a$. The figure below summarizes all these conditions.

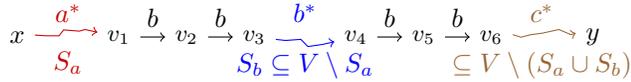

These conditions can clearly be verified in time polynomial in $G$. We develop in this paper the general idea underlying this argument, which allows us to characterize tractable instances for the problem of finding a simple path.

## 3. A TRICHOTOMY FOR RSPQ

We next define a class of languages. We will prove that it is exactly the class of regular languages for which RSPQ$(L)$ is tractable.

DEFINITION 1. *For each $i \geq 0$, we define $trC(i)$ as the class of regular languages $L$ such that for all words $w_l, w_m, w_r \in \Sigma^*$ and all non empty words $w_1, w_2 \in \Sigma^+$, if $w_l w_1^i w_m w_2^i w_r \in L$ then $w_l w_1^i w_2^i w_r \in L$.*

*We define the class $trC$ as the union $\bigcup_{i \geq 0} trC(i)$.*

LEMMA 1. *$trC$ is closed by intersection, union and word reversing.*

This result is trivial by definition. The next lemma states that we need not consider $trC(i)$ for $i > M$.

LEMMA 2. *The two statements hold:*

- *For every $i$, $trC(i) \subseteq trC(i+1)$.*
- *Let $L$ be a regular language and $M$ the size of $Q_L$. $L \in trC$ iff $L \in trC(M)$.*

PROOF. 1) Let $L \in trC(i)$ and let $w$ be a word from $L$ of the form $w = w_l w_1^{i+1} w_m w_2^{i+1} w_r$. Then $w_l w_1^{i+1} w_2^{i+1} w_r$ also belongs to $L$ since $w$ can be decomposed as $w = (w_l w_1) w_1^i w_m w_2^i (w_2 w_r)$.

2) Let $L \in trC$ and let $M$ be the size of $Q_L$. There exists $i \geq 0$ such that $L \in trC(i)$. We will prove that $L \in trC(M)$. The case when $i \leq M$ is a consequence of the previous statement. Assume that $i > M$. Let $w = w_l w_1^M w_m w_2^M w_r \in L$. By the Pumping Lemma for finite automata, there are $k, k' > 0$ such that for every $j, j' \geq 0$, $w_l w_1^{M+kj} w_m w_2^{M+k'j} w_r \in L$. Thus $w_l w_1^j w_2^{j'} w_r \in L$ for every $j, j' \geq i$, since $L \in trC(i)$. Applying once more the pumping lemma gives $w_l w_1^M w_2^M w_r \in L$. □

Here and henceforth, $M$ refers to the size of $Q_L$. We now give a characterization of $trC$ in terms of automata.

LEMMA 3. *Let $L$ be a regular language. Then, $L$ belongs to $trC$ iff for every states $q_1, q_2 \in Q_L$ such that $Loop(q_1) \neq \emptyset$, $Loop(q_2) \neq \emptyset$ and $q_2 \in \Delta(q_1, \Sigma^*)$ and for every $w \in Loop(q_2)$, the following statement holds: $w^M \mathcal{L}_{q_2} \subseteq \mathcal{L}_{q_1}$.*

PROOF. (only if) Let $q_1, q_2 \in Q_L$, $w_1 \in Loop(q_1)$ and $w_2 \in Loop(q_2)$ such that $q_2 \in \Delta(q_1, \Sigma^*)$. Let $w_l, w_m, w_r \in \Sigma^*$ such that $\Delta(i_L, w_l) = q_1$, $\Delta(q_1, w_m) = q_2$ and $w_r \in \mathcal{L}_{q_2}$. Thus, $w_l w_1^M w_m w_2^M w_r \in L$. By definition of $trC$, $w_l w_1^M w_2^M w_r \in L$ and, consequently, $w_2^M w_r \in \mathcal{L}_{q_1}$.

(if) Let $L$ be a language that satisfies the right assumption of the equivalence. We first prove that $L$ is aperiodic. Indeed, let $q$ be a state of $A_L$, $w' \in \Sigma^*$ and $k > 0$ such that $\Delta(q, w'^k) = q$. By applying the assumption, with $q_1 = q$, $q_2 = \Delta(q, w')$ and $w = w'^k$. We obtain $\mathcal{L}_{\Delta(q,w')} \subseteq \mathcal{L}_q$. Symmetrically, with $q_1 = \Delta(q, w')$, $q_2 = q$ and $w = w'^k$, we obtain $\mathcal{L}_q \subseteq \mathcal{L}_{\Delta(q,w')}$. Thus, by minimality of $A_L$, $\Delta(q, w') = q$.

Let us now prove $L \in trC(2M)$. Consider some words $w_l, w_m, w_r, w_1,$ and $w_2$ such that $w_l w_1^{2M} w_m w_2^{2M} w_r \in L$ with $w_1, w_2$ non empty. Let $q_1 = \Delta(i_L, u_l w_1^{2M})$ and $q_2 = \Delta(i_L, u_l w_1^{2M} w_m w_2^M)$. Then $w_2^M w_r \in \mathcal{L}_{q_2}$ and, since $L$ is aperiodic, $w_1 \in Loop(q_1)$ and $w_2 \in Loop(q_2)$. By hypothesis we then get $w_2^{2M} w_r \in \mathcal{L}_{q_1}$, so $w_l w_1^{2M} w_2^{2M} w_r \in L$. □

### 3.1 Hard languages for RSPQ

This section is devoted to the proof of a hardness result: RSPQ$(L)$ is NP-hard for every regular language $L$ that does not belong to $trC$. The first step toward that proof lies in the following characterization of $trC$.

LEMMA 4. *Let $L$ be a regular language. $L \in trC$ iff $L$ does not satisfy the following property:*
(1) *there exist a state $q \in Q_L$, some word $w_r \in \Sigma^*$ and some non-empty words $w_1 \in Loop(q)$ and $w_2, w_m \in \Sigma^+$ such that*

- $w_m w_2^* w_r \subseteq \mathcal{L}_q$
- $(w_1 + w_2)^* w_r \cap \mathcal{L}_q = \emptyset$

The "only if" direction is trivial. To prove the "if" implication, we assume that $L$ does not satisfy Property (1) and show the following two claims.

CLAIM 1. *Let $q_1, q_2 \in Q_L$ such that $q_2 \in \Delta(q_1, \Sigma^*)$. Then, either $Loop(q_1) \cap Loop(q_2) = \emptyset$ or $\mathscr{L}_{q_2} \subseteq \mathscr{L}_{q_1}$.*

PROOF OF CLAIM 1. Let $q_1$ and $q_2$ that do not satisfy the claim. Then, we choose $w_1 = w_2 \in Loop(q_1) \cap Loop(q_2)$ and $w_r \in \mathscr{L}_{q_2} \setminus \mathscr{L}_{q_1}$. Furthermore, we choose $w_m$ such that $\Delta(q_1, w_m) = q_2$. Thus, Property (1) holds. □

CLAIM 2. *$L$ is aperiodic.*

PROOF OF CLAIM 2. Let $q_1 \in Q_L$ and $k \geq 0$ such that $\Delta_L(q_1, w^k) = q_1$. Let $q_2 = \Delta(q_1, w)$. By Claim 1, $\mathscr{L}_{q_1} = \mathscr{L}_{q_2}$ and thus, by minimality of $A_L$, $q_1 = q_2$. □

We can now prove Lemma 4.

PROOF. Fix a language $L \notin trC$. By Lemma 3 there exist $q, q_2, w_1, w_2, w_m, w_r$ such that $w_1 \in Loop(q)$, $w_2 \in Loop(q_2)$, $\Delta(q, w_m) = q_2$, and $w_2^M \mathscr{L}_{q_2} \setminus \mathscr{L}_q \neq \emptyset$. As $L$ is aperiodic, $\Delta(q, (w_2)^M) = \Delta(q, ((w_2)^M)^M)$ hence $((w_2)^M)^M \mathscr{L}_{q_2} \setminus \mathscr{L}_q \neq \emptyset$. W.l.o.g, we can therefore suppose that $w_2 = (w_2')^M$ for some word $w_2'$, and fix some word $w_r \in w_2^M \mathscr{L}_{q_2} \setminus \mathscr{L}_q$. Consequently, every state $q'$ in $\Delta(q, \Sigma^* w_2)$ satisfies $\Delta(q', w_2) = q'$ (as $L$ is aperiodic), hence $\mathscr{L}_{q'} \subseteq \mathscr{L}_q$ by Claim 1, and consequently $(w_1 + w_2)^* w_r \cap \mathscr{L}_q$ is a subset of $w_2^* w_r \cap \mathscr{L}_q = \emptyset$. Thus, Property (1) is satisfied, which concludes the proof of Lemma 4. □

We can now prove our hardness result, by reduction from Vertex-Disjoint-Path, a problem also used in [29] to prove hardness in the particular case of $a^* b a^*$.

**Vertex-Disjoint-Path**
**Input:** A directed graph $G = (V, E)$, four vertices $x_1, y_1, x_2, y_2 \in V$
**Question:** are there two disjoint paths, one from $x_1$ to $y_1$ and the other from $x_2$ to $y_2$?

LEMMA 5. *Let $L$ be a regular language that does not belong to $trC$. Then, $RSPQ(L)$ is NP-hard.*

PROOF. As explained, we construct a reduction from the Vertex-Disjoint-Path problem. Let $q, w_m, w_r, w_1, w_2$ be defined as in Property (1) and $w_l$ such that $\Delta(i_L, w_l) = q$. We build from $G$ the db-graph $G'$.

We consider here db-graphs where edges are labeled by non empty words. This is actually a generalization of db-graphs. Nevertheless, by adding intermediate vertices, an edge labeled by a word $w$ can be replaced with a path whose edges form the word $w$.

$G'$ is constructed as follows. We start from an empty graph $G'$ whose vertices are vertices of $G$. For each edge $(v_1, v_2)$ in $G$, we add two edges $(v_1, w_1, v_2)$ and $(v_1, w_2, v_2)$. Moreover, we add two new vertices $x, y$ and three edges $(x, w_l, x_1)$, $(y_1, w_m, x_2)$ and $(y_2, w_r, y)$. Note that every simple path from $x$ to $y$ in $G'$ matches a word in $w_l(w_1+w_2)^* w_r$ or $w_l(w_1+w_2)^* w_m(w_1+w_2)^* w_r$.

Thus, $RSPQ(L)$ returns True for $(G', x, y)$ iff there is a simple path from $x$ to $y$ in $G'$ that contains the edge $(y_1, w_m, x_2)$ that is, iff Vertex-Disjoint-Path returns True for $(G, x_1, y_1, x_2, y_2)$. We illustrate below the reduction on an instance $(G, x_1, y_1, x_2, y_2)$ for $L = a^* b(cc)^* d$, choosing $w_l = w_1 = a$, $w_m = b$, $w_2 = cc$, and $w_r = d$. □

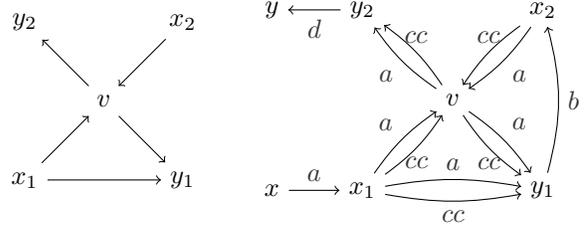

Input instance $G$   RSPQ instance: graph $G'$

**Figure 1: Reduction for $L = a^* b(cc)^* d$.**

## 3.2 Tractable languages for RSPQ

The main result of this section is that for every $L \in trC$, $RSPQ(L) \in NL$. For this purpose, we first prove several lemmas on the structure of automata that recognize $trC$ languages. In this section, we fix a language $L \in trC$. Note that $L$ satisfies Claims 1 and 2.

### 3.2.1 Technical lemmas on the components of $A_L$

Here and thereafter, we fix $N = 2M^2$. The next lemmas give information on the structure of $A_L$ components. The first lemma strengthens Lemma 3.

LEMMA 6. *Let $L$ be a regular language. Then, $L$ belongs to $trC$ iff for every states $q_1, q_2 \in Q_L$ such that $Loop(q_1) \neq \emptyset$, $Loop(q_2) \neq \emptyset$ and $q_2 \in \Delta(q_1, \Sigma^*)$, the following statement holds: $(Loop(q_2))^M \mathscr{L}_{q_2} \subseteq \mathscr{L}_{q_1}$.*

PROOF. $2 \Rightarrow 1$ is trivial.
$1 \Rightarrow 2$: Let $q_1$ and $q_2$ two states such that $Loop(q_1) \neq \emptyset$ and $q_2 \in \Delta(q_1, \Sigma^*)$. Let $v_1, \ldots v_M \in (Loop(q_2))^M$ and $q_3 = \Delta(q_1, v_1 \ldots v_M)$. We wish to prove $\mathscr{L}_{q_2} \subseteq \mathscr{L}_{q_3}$.

For some $i, j$, $1 \leq i < j \leq M$, we get $\Delta(q_1, v_1 \ldots v_i) = \Delta(q_1, v_1 \ldots v_j)$ (using the convention $\Delta(q_1, v_1 \ldots v_i) = q_1$ for $i = 0$). Let $u_1 = v_1 \ldots v_i$, $u_2 = v_{i+1} \ldots v_j$ and $u_3 = v_{j+1} \ldots v_k$. Let $q_4 = \Delta(q_1, u_1)$. Since $\mathscr{L}_{q_2} = u_3^{-1} \mathscr{L}_{q_2}$ and $\mathscr{L}_{q_3} = u_3^{-1} \mathscr{L}_{q_4}$, it suffices to prove that $\mathscr{L}_{q_2} \subseteq \mathscr{L}_{q_4}$. Let $w = u_1 u_2^M$ and $q_5 = \Delta(q_1, w^M)$. Both $u_2$ and $w$ belong to $Loop(q_5)$ because $L$ is aperiodic. As $\Delta(q_1, w^M) = q_5$ and $w \in Loop(q_2)$, we get $\mathscr{L}_{q_2} \subseteq \mathscr{L}_{q_5}$ through Lemma 3 with $q_1, q_2$ and $w$. As $\Delta(q_4, u_2^M) = q_4$ and $u_2 \in Loop(q_5)$, one more application of Lemma 3 with $q_2, q_4$ and $u_2$ yields $\mathscr{L}_{q_5} \subseteq \mathscr{L}_{q_4}$. □

This implies that two distinct states $q_1$ and $q_2$ in a same component cannot loop on a same word.

LEMMA 7. *Let $L$ be a regular language in $trC$. Let $q_1, q_2$ two states belonging to the same component of $A_L$. Then, $Loop(q_1) \cap Loop(q_2) \neq \emptyset$ implies $q_1 = q_2$.*

PROOF. Let $q_1, q_2$ as above, and let $w$ a word in $Loop(q_1) \cap Loop(q_2)$. According to Lemma 6, $w^M \mathscr{L}_{q_2} \subseteq \mathscr{L}_{q_1}$, hence $\mathscr{L}_{q_2} \subseteq \mathscr{L}_{q_1}$ since $w \in Loop(q_1)$. By symmetry, $\mathscr{L}_{q_2} = \mathscr{L}_{q_1}$, which implies $q_2 = q_1$. □

LEMMA 8. *Let $C$ be a component of $A_L$, $q_1, q_2 \in C$ and $a \in \Sigma$. Then $\Delta_L(q_1, a) \in C$ iff $\Delta_L(q_2, a) \in C$.*

PROOF. Let $q_1 \neq q_2$ two states in the same component. Assume by contradiction that $\Delta_L(q_1, a) \in C$ and $\Delta_L(q_2, a) \notin C$. Notice that $Loop(q_1)$ and $Loop(q_2)$ are not empty. Let $w \in Loop(q_1) \cap a\Sigma^*$. Let $q_3 = \Delta_L(q_2, w^M)$. As $L$ is aperiodic, $\Delta_L(q_3, w^M) = q_3$. Thus, $w^M \in Loop(q_1) \cap Loop(q_3)$. Consequently, $w^M \mathscr{L}_{q_3} \subseteq \mathscr{L}_{q_1}$ and $w^M \mathscr{L}_{q_1} \subseteq \mathscr{L}_{q_3}$. $\mathscr{L}_{q_3} \subseteq \mathscr{L}_{q_1}$ and $\mathscr{L}_{q_1} \subseteq \mathscr{L}_{q_3}$. Thus, $\mathscr{L}_{q_1} = \mathscr{L}_{q_3}$ and, by minimality of $A_L$, $q_1 = q_3$. That is an absurd because $q_1$ and $q_3$ are not in the same component. □

NOTATION 1. *We define the internal alphabet of a component $C$ of $A_L$ as the set $\Sigma_C = \{a \in \Sigma : \exists q_1, q_2 \in C . \Delta_L(q_1, a) = q_2\}$.*

As a direct consequence of Lemma 8 we get:

LEMMA 9. *Let $C$ be a component of $A_L$, $q \in C$ and $w \in \Sigma^*$. Then $\Delta(q, w) \in C$ iff $w \in (\Sigma_C)^*$.*

LEMMA 10. *Let $C$ be a component of $A_L$, $\Sigma_C$ be the internal alphabet of $C$, $q_1, q_2$ be two states of $C$ and $w \in (\Sigma_C)^{M^2}$. Then, $\Delta_L(q_1, w) = \Delta_L(q_2, w)$.*

PROOF. Assume that $w = a_1 \ldots a_{M^2}$. For each $i = 0, \ldots, M^2$ and $\alpha = 1, 2$, let $q_{\alpha,i} = \Delta_L(q_\alpha, a_1 \ldots a_i)$. Since there are at most $M^2$ distinct pairs $(q_{1,i}, q_{2,i})$, there exist $i, j$, with $i < j$ such that $q_{1,i} = q_{1,j}$ and $q_{2,i} = q_{2,j}$. Let $w' = a_{i+1} \ldots a_j$. We have $w' \in Loop(q_{1,i}) \cap Loop(q_{2,i})$. Thus, by Lemma 9, $q_{1,i}, q_{2,i} \in C$ and, by Lemma 7, $q_{1,i} = q_{2,i}$ and $\Delta_L(q_1, w) = \Delta_L(q_2, w)$. □

LEMMA 11. *Let $q_1, q_2$ be two states such that $q_2 \in \Delta_L(q_1, \Sigma^*)$, $Loop(q_1) \neq \emptyset$, and $Loop(q_2) \neq \emptyset$. Let $C$ be the component that contains $q_2$ and $\Sigma_C$ be the internal alphabet of $C$. Then, $\mathscr{L}_{q_2} \cap (\Sigma_C)^N \Sigma^* \subseteq \mathscr{L}_{q_1}$.*

PROOF. Let $w \in \mathscr{L}_{q_2} \cap (\Sigma_C)^N \Sigma^*$. There are some words $u, v \in (\Sigma_C)^{M^2}$, $w' \in \Sigma^*$ such that $w = uvw'$. By Lemma 8 and the Pigeonhole Principle, there exist a state $q_3$ and $M+1$ non-empty words $v_1, \ldots, v_{M+1}$ such that $v = v_1 \ldots v_{M+1}$ and $\Delta_L(q_2, uv_1 \ldots v_i) = q_3$ for every $i \in [M]$. Therefore, $w \in uv_1(Loop(q_3))^{M-1}v_{M+1}w'$. By Lemma 10, $\Delta_L(q_3, uv_1) = \Delta_L(q_2, uv_1) = q_3$, Thus, $w \in Loop(q_3)^M v_{M+1} w' \cap \mathscr{L}_{q_3}$. By Lemma 3, $w \in \mathscr{L}_{q_1}$. □

### 3.2.2 Computing $\mathrm{RSPQ}(L)$ for $L$ in $trC$

In the following, we describe a polynomial algorithm that computes $\mathrm{RSPQ}(L)$ when $L$ belongs $trC$. Observe that if we are looking for a (non necessarily simple) regular path, a dynamic programming approach can be used, essentially because only the last vertex in the (partial) path needs to be memorized in order to build the path incrementally. This approach is not adequate to build a simple path, as we need to memorize all the vertices in the path. We therefore need to consider an exponential number of paths.

Nevertheless, we will show that in the case where $L$ belongs to $trC$, we can identify a finite number of vertices that suffice to check if the path is (or can be transformed into) a simple path labeled with $L$. These "important" vertices shall be stored in a path summary, as presented in the following. Unlike paths, summaries can be enumerated in logarithmic space, and we shall explain how one can use the summaries to check if there exists a simple path between the input nodes.

We first define the notion of $L$-annotation of a path $p$.

DEFINITION 2. *Let $L$ be a language in $trC$, and let $p$ be a path $p = (v_1, a_1, \ldots, a_m, v_{m+1})$. the $L$-annotation of $p$ is the mapping $\rho : \{v_1, \ldots, v_{m+1}\} \mapsto Q_L$ such that: $\rho(v_1) = i_L$ and $\rho(v_{i+1}) = \Delta_L(i_L, a_1 \ldots a_i)$ for every $i \in [m]$.*

We now introduce the concept of summary for a path $p$ (with annotation $\rho$). Roughly speaking, the idea is to keep only a bounded number of vertices of $p$ (that depends only on $L$). Actually, the information we must record for each component $C$ of $A_L$ can be limited to the first and the $N$ last vertices having their state in $C$. This allows us to apply Lemma 11. Additionally, if the number of such vertices is greater than $N+1$, we replace the path between the first vertex and the $N$ last ones by a special symbol $\Sigma_C^*$ where $\Sigma_C$ is the internal alphabet of $C$. It means that the path we have removed forms a word that belongs to $\Sigma_C^*$. More formally, a summary is defined as follow.

DEFINITION 3. *Let $p = (v_1, a_1, \ldots, a_k, v_{m+1})$ be a path and let $\rho$ the $L$-annotation of $p$. The summary $S$ of the path $(p, \rho)$ (w.r.t. $A_L$) is obtained from $p$ (and $\rho$) by the following replacements. Let $C_i, \ldots, C_l$ be the components such that there are at least $N+1$ vertices $v$ in $p$ such that $\rho(v) \in C_i$ (the sequence is sorted in topological order). For each component $i \in [l]$, let $\alpha_i$ and $\beta_i$ denote the first and the maximal indices such that $\rho(v_{\alpha_i}), \rho(v_{\beta_i}) \in C_i$. Let $\beta_i' = \beta_i - N$. We replace the subpath $v_{\alpha_i} \ldots v_{\beta_i'}$ by $v_{\alpha_i}, \Sigma_C^*, v_{\beta_i'}$. We denote by $lrc(p)$ the set $\{C_i, \ldots, C_l\}$ of long run components.*

Notice that a summary contains at most $NM = 2M^3$ elements (vertices and labels), which is constant if $L$ is fixed. Consequently, given a db-graph $G$, the number of distinct summaries of $L$-labeled paths in $G$ is bounded by a polynomial in $|G|$ (when $L$ is fixed). Furthermore, each vertex of the graph can be represented with a logarithmic number of bits. A logarithmic number of bits is therefore sufficient to encode a summary (for fixed $L$).

We define a candidate summary $S$ as a sequence of vertices and labels of the form above; $S = (v_1, \alpha_1, \ldots, \alpha_l, v_{l+1})$ where $\alpha_i \in \Sigma \cup \{\Sigma_C^* \mid C\}$ and $l \leq NM$. A path $p$ obtained by replacing each subsequence $(v_{\alpha_i}, \Sigma_{C_i}^*, v_{\beta_i'})$ with a simple $\Sigma_{C_i}^*$-labeled path from $v_{\alpha_i}$ to $v_{\beta_i'}$ is called a *completion* of the candidate summary $S$.

EXAMPLE 2. *Figure 2 represents the minimal DFA for $L = a(c^{\geq 2} + \epsilon)(a+b)^*(ac)?a^*$ (we did not represent the sink state). This automaton can loop in three strongly connected components: $C_1 = \{q_4\}$, $C_2 = \{q_5, q_6\}$, and $C_3 = \{q_7\}$. The accepting states are $q_2$, $q_4$, $q_5$, $q_6$, and $q_7$. In this automaton $M = 7$, so $N$ should be huge but we shall pretend that $N = 3$ for our example as this value is sufficient for our algorithm. Let us consider the path $p_1$ illustrated in Figure 3 with thick edges. The table below details this path and the corresponding annotation.*

*We observe that $p_1$ is a simple $L$-labeled path. The*

| $v_1$ | $v_2$ | $v_3$ | $v_4$ | $v_5$ | $v_7$ | $v_8$ | $v_9$ | $v_{10}$ | $v_{11}$ | $v_{12}$ | $v_{13}$ | $v_{14}$ | $v_{15}$ |
|---|---|---|---|---|---|---|---|---|---|---|---|---|---|
| $q_1$ | $q_2$ | $q_3$ | $q_4$ | $q_4$ | $q_4$ | $q_4$ | $q_4$ | $q_5$ | $q_6$ | $q_5$ | $q_5$ | $q_5$ | $q_5$ |
| | | | annotations in $C_1$ | | | | | annotations in $C_2$ | | | | | |

*summary $S$ of $p_1$ is obtained by removing the second (resp. second and third) vertex with state annotated in $C_1$ (resp. $C_2$): only the vertices highlighted in red in the table are preserved, and the component for the vertices eliminated is identified by special symbols $\Sigma_{C_1}^*$ and $\Sigma_{C_2}^*$.*

$$S = (v_1, a, v_2, c, v_3, c, v_4, \Sigma_{C_1}^*, v_7, c, v_8, c, v_9,$$
$$a, v_{10}, \Sigma_{C_2}^*, v_{13}, a, v_{14}, a, v_{15}).$$

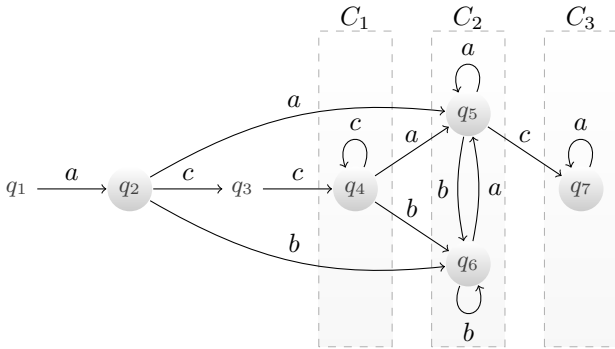

**Figure 2:** Minimal DFA for $a(c^{\geq 2}+\epsilon)(a+b)^*(ac)?a^*$

LEMMA 12. *Let $S$ be the summary of an $L$-labeled path $p$ and let $p'$ be a completion of $S$. Then, $p'$ is an $L$-labeled path of summary $S$.*

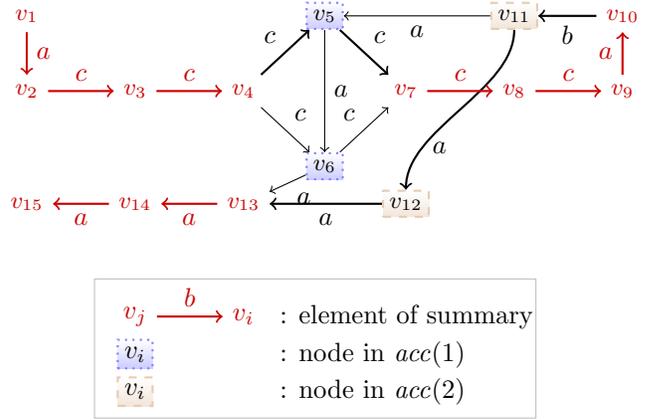

| $v_j \xrightarrow{b} v_i$ | : element of summary |
|---|---|
| $v_i$ (dashed blue) | : node in $acc(1)$ |
| $\bar{v_i}$ (dashed orange) | : node in $acc(2)$ |

**Figure 3:** Nice simple path, and its summary.

This gives an NL algorithm to test if a given candidate summary $S$ is the summary of an $L$-labeled path: we only need to compute a completion and then test if this completion is an $L$-labeled path. However, the completed path is not necessarily simple, even if $S$ is the summary of a simple path. Indeed, the paths we have built between each $v_{\alpha_i}$ and $v_{\beta_i'}$ are not necessarily disjoint. To overcome this problem, we will restrict the set of potential paths and we will only search a certain type of path that we name nice path.

DEFINITION 4. *Let $p$ be an $L$-labeled path of run $\rho$ and summary $S$. Let $F$ be the set of vertices appearing in $S$. Let $(C_i, \alpha_i, \beta_i', \beta_i)_{i \in [l]}$ as stated in Definition 3. We define $P_i$, $\text{length}_i$ and $acc(i)$ as follows for every $i \in [l]$ in increasing order: $P_i$ is the set of paths $p'$ starting from $v_{\alpha_i}$ and satisfying the following three conditions:*

1. *$p'$ is a simple $\Sigma_{C_i}^*$-labeled path;*
2. *there is no vertex in $p'$ that belongs to $F \setminus \{v_{\alpha_i}, v_{\beta_i'}\}$*
3. *furthermore, there is no vertex in $p'$ that belongs to $acc(j)$ for any $j \in [i-1]$.*

*We define $\text{length}_i$ as the length of the shortest path from $v_{\alpha_i}$ to $v_{\beta_i'}$ that belongs to $P_i$. And $acc(i)$ is the set of vertices $y$ reachable from $v_{\alpha_i}$ by a path $p \in P_i$ of size $w(p) \leq \text{length}_i$.*

*We qualify $p$ as* nice *if the following three conditions are satisfied: (a) all vertices appearing in $S$ are distinct and (b) the subpath $p_i$ of $p$ from $v_{\alpha_i+1}$ to $v_{\beta_i'}$ belongs to $P_i$ and (c) $w(p_i) = \text{length}_i$. Consequently all vertices of $p_i$ belong to $acc(i)$.*

EXAMPLE 3. *The path $p_1$ defined in Example 2 is nice, since $acc(1) = \{v_5, v_6\}$ and $acc(2) = \{v_{11}, v_{12}\}$, as illustrated in Figure 3. Observe that neither $v_5$ nor $v_6$ are necessary to fill the gap from $v_4$ to $v_7$ in the summary, yet one of them is necessarily traversed. The definition of nice paths guarantees the paths $p_1''$ and $p_2''$ replacing $\Sigma_{C_1}$ and $\Sigma_{C_2}$ are disjoint.*

One might think the intersection of the two paths $p_1''$ and $p_2''$ does not matter, as each loop can be eliminated while preserving the membership of the label in $L$, according to Lemma 11. In this example this is indeed the case: the path through $v_1, v_2, v_3, v_4, v_6, v_{13}, v_{14}$, and $v_{15}$ belongs to the language $L$. But Example 4 shows this needs not be the case in general.

EXAMPLE 4. *The language $L = a^*(bb^+ + \varepsilon)c^*$ from Example 1 shows why one cannot simply eliminate loops by repeated applications of Lemma 11. The minimal DFA for $L$ has 4 states so $N$ should be $2*16$ but here we could assume $N = 2$. The $L$-labeled path illustrated in Figure 4 consists of an $a^{2N}$ path from $x_0$ to $x_{2k}$, a $c^{2N}$-labeled path $p_y$ from $y_0$ to $y_{2k}$, plus a $b^{2N}$-labeled path from $x_{2k}$ to $y_0$, intersecting $p_x$ and $p_y$ in their middle after $N$ and $N+1$ letters, respectively. This path is not simple as it intersects twice with itself. Lemma 11 can be applied once to remove any of the two loops but then the remaining loop cannot be removed while preserving the language. The notion of nice path has been introduced to tackle exactly this difficulty.*

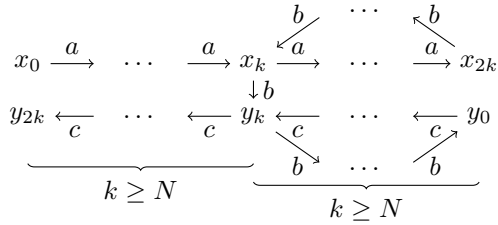

**Figure 4: Counterexample to loop-elimination.**

LEMMA 13. *Let $p$ be an $L$-labeled path from $x$ to $y$. If $p$ is nice, then it is a simple $L$-labeled path.*

PROOF. This is because for any nice path the nodes of its summary are distinct, the path inside a component is simple due to condition 1, and the sets $acc(i)$ are pairwise disjoint due to condition 3. □

We observe that, although condition 1 requires that the path inside a component are simple, we can easily obtain a nice path from any $p$ that violates no condition but 1 by removing the loops inside a component. The whole point of the definition is to make sure that there are no loops in the path except inside a connected component.

The following lemma explains that finding a simple path is equivalent to finding a nice path.

LEMMA 14. *Let $(G, x, y)$ an instance of $RSPQ(L)$. Then, every shortest simple $L$-labeled path from $x$ to $y$ is nice.*

PROOF. Let $p = (v_1, a_1, \ldots, a_m, v_{m+1})$ be a shortest simple $L$-labeled path from $x$ to $y$. We use the notations $(C_i, \alpha_i, \beta_i', \beta_i)_{i \in [l]}$ as stated in Definition 3 for the long run components in the summary $S$ of $p$ and their corresponding indices.

Assume for the sake of contradiction that $p$ is not nice. There exists $i \in [l], j \in [l], k \in [m]$ such that $i < j$, $\alpha_i < k < \beta_i'$, and $v_k \in acc(i)$. We choose $i$ as minimal and then $k$ as maximal for this property. Let $p_1$ be the subpath of $p$ from $v_{\alpha_i}$ to $v_k$. By definition there is a $\Sigma_{C_i}^*$-labeled simple path $p_2$ from $v_{\alpha_i}$ to $v_k$ whose vertices belong to $acc(i)$. We define $p'$ from $p$ by replacing the subpath $p_1$ by $p_2$. Let $\rho'$ be the $L$-annotation of $p'$. Notice that $\rho(v_k) \in C_j$ and $\rho'(v_k) \in C_i$. Furthermore, $a_k \ldots a_m \in \Sigma_{C_j}^N \Sigma^* \cap \mathscr{L}_{\rho(v_k)}$. Thus, by Lemma 11, $a_k \ldots a_m \in \mathscr{L}_{\rho'(v_k)}$. Consequently, $p'$ is a $L$-labeled path.

We now prove that $p'$ is simple. Since $p$ is simple, it suffices to prove that the vertices in $p_2$ are disjoint with other vertices in $p'$. By minimality of $i$, the sets $acc(1), \ldots acc(i)$ are pairwise disjoint. By maximality of $k$, for every $k' > k$, $v_j \notin acc(i)$.

We finally observe that $p'$ is strictly shorter than $p$ since $w(p_2) < w(p_1)$. It is in contradiction with the minimality of $w(p)$. □

We next show how a nice summary can be completed in logarithmic space into a simple path.

LEMMA 15. *Let $L$ be a fixed language in $trC$. There exists a non deterministic log-space algorithm that given an instance $(G, x, y)$ of $RSPQ(L)$ and a candidate summary $S$*

  1. *returns "Yes" if there is a shortest simple $L$-labeled path of summary $S$ from $x$ to $y$;*

  2. *returns "No" if there is no simple $L$-labeled path from $x$ to $y$;*

  3. *is unspecified otherwise.*

PROOF. It suffices to complete the summary $S$ into a *nice* path $p$. If $S$ is a summary of a nice $L$-labeled path then $p$ is a $L$-labeled path by Lemma 12 and is simple by Lemma 13. If such path do not exist, the algorithm returns "No". This can be done in logarithmic space by directly applying the definition. To this purpose, we need to compute the sets $acc(i)$. These sets $acc(i)$ can be described in FO + TC i.e. FO plus the transitive closure operator. Since the evaluation of a fixed FO + TC formula is in NL [20], the sets $acc(i)$ can be computed using a logarithmic space. Notice that the sets $acc(i)$ are not stored in memory but computed each time we need to access them. The same remark applies to the path $p$. The candidate summary $S$ may not be a summary of a simple $L$-labeled path. Thus, we must check that the obtained path is a simple $L$-labeled path. Notice that the algorithm possibly returns "Yes" even when $S$ is not a summary of a shortest path. □

We eventually show the main Lemma of this section.

LEMMA 16. *Let $L \in trC$. Then, $RSPQ(L) \in$ NL.*

PROOF. We simply enumerate all possible candidate summaries $S$ w.r.t. $(L, G, x, y)$, and apply on each summary the algorithm of Lemma 15. We return "Yes" if this algorithm returns "Yes" for at least one candidate summary $S$. Otherwise, we return "No". Therefore, our algorithm returns "Yes" if and only if there exists a nice path from $x$ to $y$, and consequently, if and only if there is a simple path from $x$ to $y$. Since $L$ is fixed, there is a polynomial number of candidate summaries, each of logarithmic size. Consequently, they can be enumerated within logarithmic space. □

Notice that we can easily adapt our algorithm such that it outputs a shortest path for positive instances. It can be generalized to db-graphs weighted by a function $E \to \mathbb{R}^+$.

We can now state the main theorem.

THEOREM 1. *Let $L$ be a regular language. Then, $RSPQ(L)$ is in NL if $L \in trC$ and is NP-complete otherwise.*

## 3.3 Towards a complete classification

Actually, the classification can be made more precise. We have divided the $RSPQ(L)$ problems into NL and NP-complete problems. Now, we can envisage a classification within the class of NL problems.

LEMMA 17. *For every regular language $L$, $RSPQ(L) \in$ $AC^0$ if $L$ is finite and is NL-hard otherwise.*

The proof is based on a reduction from the following NL-complete problem [33]

**Reachability**
**Input:** A directed graph $G$ and two vertices $x, y$ in $G$
**Question:** is there a path from $x$ to $y$?

PROOF. (Easiness) Given an alphabet $\Sigma$, we consider the signature $\tau = (R_a)_{a \in \Sigma})$ of binary predicates. We can view a db-graph $(V, \Sigma, E)$ as a $\tau$-structure $\mathcal{M} = (V, (R_a)_{a \in \Sigma})$ of domain $V$ and such that $(v_1, v_2) \in R_a$ iff $(v_1, a, v_2) \in E$ for every $v_1, v_2 \in V$ and $a \in \Sigma$. Let $w = a_1 \ldots a_k$ be a word. We let the reader verify that the predicate $\text{path}_w(x, y)$ (which means there is a simple $w$-labeled path between $x$ and $y$) is expressible in FO.

(Hardness) We make a a reduction from Reachability. Let $L$ be an infinite regular language. By the Pumping Lemma, there exist non empty words $u, v, w$ such that the language $uv^*w \subseteq L$. We build an db-graph $G'$ from $G$ by labeling each edge of $G$ with $v$. We add two vertices $x'$ and $y'$ and two edges $(x', u, x)$ and $(y, w, y')$.

There is a (simple) path from $x$ to $y$ in $G$ iff there is a $L$-labeled simple path from $x$ to $y$. Consequently, $RSPQ(L)$ is NL-hard. □

Our results so far can now be summarized by the following trichotomy which refines Theorem 1.

THEOREM 2. *Let $L$ be a regular language. One of these statements hold.*

1. *$L$ is finite: $RSPQ(L) \in AC^0$;*
2. *$L \in trC$ and $L$ is infinite: $RSPQ(L)$ is NL-complete;*
3. *$L \notin trC$: $RSPQ(L)$ is NP-complete.*

## 3.4 Recognition of tractable languages

This section investigates the complexity of deciding if $RSPQ(L)$ is tractable (i.e. if $RSPQ(L)$ can be computed in polynomial time). We consider different representations of $L$ (DFAs, NFAs or regular expressions).

THEOREM 3. *Testing whether a regular language $L$ belongs to $trC$ is:*

1. *NL-complete if $L$ is given by a DFA;*
2. *PSPACE-complete if $L$ is given by an NFA (resp. a regular expression).*

The proofs of hardness for DFA and NFA rely on reductions from the following two problems respectively.

**Emptiness**
**Input:** A DFA $A_L = (Q_L, \Sigma, i_L, F_L, \Delta_L)$ that recognizes a language $L$
**Question:** is $L = \emptyset$ ?

and

**Universality**
**Input:** An NFA (or a regular expression) that recognizes a language $L \subseteq \{0,1\}^*$
**Question:** $L = \{0,1\}^*$ ?

The NL-completeness of Emptiness can easily be deduced from the NL-completeness of Reachability [33]. Stockmeyer and Meyer [38] prove that Universality is PSPACE-complete.

## 3.5 Characterization by regular expressions

In this section, we propose a characterization of $trC$ languages in terms of regular expressions: the languages in $trC$ are exactly those that can be expressed with an expression in the fragment $\Psi_{tr}$ defined below. This fragment essentially enforces restrictions on the concatenation of subexpressions: except at the highest level, only expressions of the form $e + \varepsilon$ can be concatenated.

$\Psi_{tr}$-terms are defined as follows:

$$\Psi_{tr}\text{-term} ::= w + \epsilon \mid A^{\geq k} + \epsilon$$

where $w$ is a word and $A$ is a subset of $\Sigma$ and $A^{\geq k}$ is a shortcut for $A^k A^*$. A $\Psi_{tr}$-sequence is a concatenation of terms $w\varphi_1 \ldots \varphi_l w'$ where $w$ and $w'$ are words and $\varphi_1 \ldots \varphi_l$ are $\Psi_{tr}$-terms. Finally, the fragment $\Psi_{tr}$ is the set of disjunctions of $\Psi_{tr}$-sequences.

THEOREM 4. *A language $L$ belongs to $trC$ iff $L$ is recognized by a regular expression in $\Psi_{tr}$.*

PROOF. We leave the details for the Appendix. Proving that expressions in $\Psi_{tr}$ define tractable languages is relatively straightforward. Lemmas 10 and 11 are the cornerstones that allow to construct the expressions for the other direction. □

We observe that adapting the notion of summary allows for a proof of Lemma 16 that directly considers regular expressions in $\Psi_{tr}$. Since $trC$ is closed by union, we can restrict ourselves to $\Psi_{tr}$-sequences $\varphi_1 \ldots \varphi_l$ where $\varphi_1, \varphi_l$ are words and $\varphi_2, \ldots, \varphi_{l-1}$ are $\Psi_{tr}$-terms. Let $p$ be a $L$-path. We decompose $p$ into subpaths $p_1, \ldots p_l$ such that $p_i$ matches the expression $\varphi_i$ for every $i \in [l]$. The summary of $p$ is built as follows:

- if $\varphi_i$ is a word or an expression of the form $w + \epsilon$, we keep all vertices of $p_i$ in the summary;

- if $\varphi_i$ is an expression of the form $A^{\geq k} + \epsilon$, we keep the $k$ first and $k$ last vertices and replace the rest of the path by the symbol $A^*$.

## 4. OTHER RESULTS

This section investigates three further issues. First, we consider RSPQs over vertex-labeled graphs. Then, we give minor results on the parametrized complexity of the RSPQ problem, Finally, we discusses the complexity of RSPQs over graphs of bounded directed treewidth. These are straightforward applications of standard techniques, yet the results may be of practical interest.

### 4.1 Other models of database graphs

The goal of this section is to adapt our classification to two other models of graphs: vl-graphs i.e. graphs whose vertices are labeled, and evl-graphs where both vertices and edges are labeled. We denote by vlg the class of vl-graphs and evlg the class of evl-graphs.

For simplicity, we will consider vl-graphs and evl-graphs as special db-graphs. This will let us work with unique model and definitions. For vl-graphs, we can put the label of a vertex into edges. Consequently, we see a vl-graph as a db-graph that respects the following restriction: there exists no pair of edges $e = (x, a, y)$ and $e' = (x', a', y)$ such that $a \neq a'$. Similarly, a vlc-graph can be seen as a db-graph over an alphabet $\Sigma_V \times \Sigma_E$ where $\Sigma_V$ is the set of vertices labels and $\Sigma_E$ is the set of edges labels.

Clearly, given a language $L$, RSPQ$(L, \text{vlg})$ is at most as difficult as RSPQ$(L)$. However, for some languages, the problem is easier. For example, for $L = a^*bc^*$, RSPQ$(L, \text{vlg}) \in$ PTIME while RSPQ$(L)$ is NP-complete. The key is that a vertex cannot have two different labels, and, consequently, a path that matches $a^*$ is always disjoint from a path that matches $c^*$. By contrast, for $L = a^*ba^*$ or $L = (aa)^*$, the problem remains NP-complete.

By generalizing this, we can define the class $trC_{vlg}$ that is the equivalent of $trC$ for RSPQ$_{vlg}$. The idea is that we can restrict the definition to consider only words $w_1$ and $w_2$, whose last letter is identical.

NOTATION 2. *We define the relation $\equiv_{vl}$ as follows: $w_1 \equiv_{vl} w_2$ if there exists a label $a \in \Sigma$ such that $w_1, w_2 \in \Sigma^* a$. For every label $a \in \Sigma$ and state $q \in Q_L$, we define $Loop_a(q) = Loop(q) \cap \Sigma a^*$.*

DEFINITION 5. *For each $i \geq 0$, we define $trC_{vlg}(i)$ as the class of regular languages $L$ that satisfy the following condition for every words $w_1, w_2, w_l, w_m, w_r \in \Sigma^*$ such that $w_1 \equiv_{vl} w_2$: if $w_l w_1^i w_m w_2^i w_r \in L$ then $w_l w_1^i w_2^i w_r \in L$. We define the class $trC_{vlg}$ as the union $\bigcup_{i \geq 0} trC_{vlg}(i)$.*

As for db-graphs, we obtain the following dichotomy.

THEOREM 5. *Let $L$ be a regular language. Then, RSPQ$_{vlg}(L)$ is in NL if $L \in trC_{vlg}$ and is NP-complete otherwise.*

Only minor changes are required from the approach for db-graphs. Proofs are not provided here but will be given in an extended version of this paper. The three main differences are the following:

- in every proof where words $w_1$ and $w_2$ appear, we consider that $w_1 \equiv_{vl} w_2$;

- instead of considering two states $q_1$ and $q_2$ such that $Loop(q_1) \neq \emptyset$ and $Loop(q_2) \neq \emptyset$, we consider two states $q_1$ and $q_2$ and a label $a$ such that $Loop_a(q_1) \neq \emptyset$ and $Loop_a(q_2) \neq \emptyset$;

- the special symbol between $v_{\alpha_i}$ and $v_{\beta'_i}$ in summaries is no longer of the form $\Sigma_C^*$, but $\lambda(v_{\alpha_i})^{-1} L_{C_i}$ where $L_{C_i}$ is the internal language of $C_i$. Similarly, the paths $P_i$ is the definition of a nice path must be $\lambda(v_{\alpha_i})^{-1} L_{C_i}$-labeled graphs.

We can obtain a similar result on evl-graphs. We define a relation $\equiv_{evl}$ over words of the alphabet $\Sigma = \Sigma_v \times \Sigma_e$. $w_1 \equiv_{evl} w_2$ if there exists $(a_v, a_e), (a_v, a'_e) \in \Sigma$ such that $w_1 \in \Sigma^*(a_v, a_e)$ and $w_2 \in \Sigma^*(a_v, a'_e)$.

DEFINITION 6. *For each $i \geq 0$, we define $trC_{evlg}(i)$ as the class of regular languages $L$ that satisfy the following condition: for every words $w_1, w_2, w_l, w_m, w_r \in \Sigma^*$ such that $w_1 \equiv_{evl} w_2$ and $w_l w_1^i w_m w_2^i w_r \in L$, it holds $w_l w_1^i w_2^i w_r \in L$. We define the class $trC_{evlg}$ as the union $\bigcup_{i \geq 0} trC_{evlg}(i)$.*

THEOREM 6. *Let $L$ be a regular language. Then, $RSPQ(L, evlg)$ is in NL if $L \in trC_{evlg}$, and is NP-complete otherwise.*

## 4.2 Parametrized complexity

The next section focuses on the parametrized complexity of the RSPQ problem.

---
**para-RSPQ($\mathcal{L}$)**
**Input:** a db-graph graph $G = (V, \Sigma, E)$,
a regular language $L \in \mathcal{L}$ given by an NFA $A_L = (Q_L, i_L, F_L, \Delta_L)$
**Parameter:** $|Q_L|$
**Question:** Is there a simple $L$-path of size at most $k$ in $G$?

---

Our initial goal was to determine the parametrized complexity para-RSPQ($trC$). Unfortunately, we could only partially reach this goal. We first address the parametrized complexity of RSPQs when the parameter is the size of the path.

---
**k-RSPQ**
**Input:** a db-graph graph $G = (V, \Sigma, E)$,
a regular language $L$ given by an NFA $A_L = (Q_L, i_L, F_L, \Delta_L)$,
two vertices $x$ and $y$ an integer $k \geq 0$
**Parameter:** $k$
**Question:** Is there a simple $L$-labeled path of size at most $k$ from $x$ to $y$ in $G$?

---

THEOREM 7. *k-RSPQ is FPT. More precisely, the problem is solvable in time $O(2^{O(k)}|A_L| \cdot |G| \cdot \log|G|)$.*

The proof is based on the Color Coding method [2]. As a consequence of this theorem we get:

COROLLARY 1. *Let $\mathcal{L}$ be the class of finite languages. Then para-RSPQ($\mathcal{L}$) $\in$ FPT.*

The finite language can be given by an acyclic NFA or a star-free regular expression.

## 4.3 Directed treewidth

Directed treewidth is a notion introduced in [23]. It generalizes many other measures such as treewidth, dag-width or Kelly-width [7, 19]. Directed treewidth measures in some sense how close a digraph is to a DAG. Johnson et al. [23] present a general method to design polynomial algorithms on graphs of bounded directed treewidth. Like most algorithms exploiting treewidth, this method leverages a dynamic programming approach on the decomposition tree. They apply this method to show that testing the existence of an hamiltonian path is polynomial on such classes of graphs. Here, we extend this result to show that the regular simple path problem is also computable in polynomial time for the same classes.

It has been observed in the literature that RSPQ has polynomial combined complexity on two interesting classes of graphs: graphs of bounded treewidth [6], and DAGs [29]. The result for DAGs is immediate indeed, as every path in a DAG is simple. The next theorem generalizes both these two results.

THEOREM 8. *Let $k \geq 0$ and $\mathcal{G}$ be a class of db-graphs with directed treewidth at most $k$. Then, $RSPQ(Reg, \mathcal{G})$ is polynomial, where Reg denotes the regular languages.*

## 5. CONCLUSION

We now pinpoint some directions for future work.

- As an extension of our work, we can consider context-free languages. It seems to be difficult.

- We have studied the regular simple path problem from the data complexity perspective. An interesting continuation of our work is to include the language in the input (combined complexity). The question is to decide given a class of language $\mathcal{L}$ whether RSPQ($\mathcal{L}$) is in P or NP-complete. Mendelzon and Wood prove that RSPQ($\mathcal{L}$) is in P for the class $\mathcal{L}$ of languages closed by subwords, which actually corresponds to $trC(0)$. By Theorem 1, $\mathcal{L} \subseteq trC$ is a necessary condition to get polynomial combined complexity. It is not a sufficient condition because the problem can be NP-complete even for a class of finite languages. We conjecture that a sufficient condition is that there exists $i \geq 0$ such that $\mathcal{L} \subseteq trC(i)$. It is not clear whether this condition is necessary.

- What becomes tractable under restrictions to the graph such as planar digraphs or undirected graphs? Notice that both disjoint paths and even path problems are polynomial in these cases [25, 30, 34, 36].

- From the parametrized complexity perspective, what is the complexity of para-RSPQ($trC$)? We conjecture that it is in FPT.

## 6. REFERENCES

[1] S. Abiteboul and V. Vianu. Regular path queries with constraints. *J. Comput. Syst. Sci.*, 58(3):428–452, 1999.
[2] N. Alon, R. Yuster, and U. Zwick. Color-Coding. *J. ACM*, 42(4):844–856, 1995.
[3] M. Arenas, S. Conca, and J. Pérez. Counting beyond a yottabyte, or how sparql 1.1 property paths will prevent adoption of the standard. In *WWW*, pages 629–638, 2012.
[4] E. M. Arkin, C. H. Papadimitriou, and M. Yannakakis. Modularity of Cycles and Paths in Graphs. *J. ACM*, 38(2):255–274, 1991.

[5] P. Barceló, L. Libkin, and J. L. Reutter. Querying graph patterns. In *PODS*, pages 199–210. ACM, 2011.

[6] C. L. Barrett, R. Jacob, and M. V. Marathe. Formal-language-constrained path problems. *SIAM Journal on Computing*, 30(3):809–837, 2000.

[7] D. Berwanger, A. Dawar, P. Hunter, S. Kreutzer, and J. Obdrzálek. The dag-width of directed graphs. *J. Comb. Theory, Ser. B*, 102(4):900–923, 2012.

[8] D. Calvanese, G. D. Giacomo, M. Lenzerini, and M. Y. Vardi. Answering regular path queries using views. In *ICDE*, pages 389–398, 2000.

[9] D. Calvanese, G. D. Giacomo, M. Lenzerini, and M. Y. Vardi. Rewriting of regular expressions and regular path queries. *J. Comput. Syst. Sci.*, 64(3):443–465, 2002.

[10] D. Calvanese, G. D. Giacomo, M. Lenzerini, and M. Y. Vardi. Reasoning on regular path queries. *SIGMOD Record*, 32(4):83–92, 2003.

[11] D. Calvanese, G. D. Giacomo, M. Lenzerini, and M. Y. Vardi. An automata-theoretic approach to regular xpath. In *DBPL*, pages 18–35, 2009.

[12] B. Courcelle. Graph rewriting: An algebraic and logic approach. In *Handbook of Theoretical Computer Science, Volume B: Formal Models and Sematics (B)*, pages 193–242. 1990.

[13] I. F. Cruz, A. O. Mendelzon, and P. T. Wood. A graphical query language supporting recursion. In *SIGMOD Conference*, pages 323–330, 1987.

[14] W. Fan, J. Li, S. Ma, N. Tang, and Y. Wu. Adding regular expressions to graph reachability and pattern queries. In *ICDE*, pages 39–50. IEEE Computer Society, 2011.

[15] J. Flum and M. Grohe. *Parameterized Complexity Theory (Texts in Theoretical Computer Science. An EATCS Series)*. 2006.

[16] M. R. Garey and D. S. Johnson. *Computers and Intractability: A Guide to the Theory of NP-Completeness*. W. H. Freeman, 1979.

[17] C. Gutierrez, C. A. Hurtado, A. O. Mendelzon, and J. Pérez. Foundations of semantic web databases. *J. Comput. Syst. Sci.*, 77(3):520–541, 2011.

[18] R. H. Güting. GraphDB: A Data Model and Query Language for Graphs in Databases. In *Proc. 20th Int. Conf. on Very Large Data Bases*, pages 297–308, 1994.

[19] P. Hunter and S. Kreutzer. Digraph measures: Kelly decompositions, games, and orderings. *Theor. Comput. Sci.*, 399(3):206–219, 2008.

[20] N. Immerman. Nondeterministic space is closed under complementation. *SIAM J. Comput.*, 17(5):935–938, 1988.

[21] N. Immerman. *Descriptive complexity*. Springer, 1999.

[22] R. Jin, H. Hong, H. Wang, N. Ruan, and Y. Xiang. Computing label-constraint reachability in graph databases. In *SIGMOD Conference*, pages 123–134, 2010.

[23] T. Johnson, N. Robertson, P. D. Seymour, and R. Thomas. Directed tree-width. *J. Comb. Theory, Ser. B*, 82(1):138–154, 2001.

[24] A. Koschmieder and U. Leser. Regular path queries on large graphs. In *SSDBM*, pages 177–194, 2012.

[25] A. S. Lapaugh and C. H. Papadimitriou. The even-path problem for graphs and digraphs. *Networks*, 14(4):507–513, 1984.

[26] U. Leser. A query language for biological networks. In *ECCB/JBI*, page 39, 2005.

[27] L. Libkin and D. Vrgoc. Regular Path Queries on Graphs with Data. In *ICDT 2012*, pages 74–85, 2012.

[28] K. Losemann and W. Martens. The complexity of evaluating path expressions in sparql. In *PODS*, pages 101–112. ACM, 2012.

[29] A. O. Mendelzon and P. T. Wood. Finding Regular Simple Paths in Graph Databases. *SIAM J. Comput.*, 24(6):1235–1258, 1995.

[30] Z. P. Nedev. Finding an Even Simple Path in a Directed Planar Graph. *SIAM J. Comput.*, 29:685–695, 1999.

[31] Z. P. Nedev and P. T. Wood. A polynomial-time algorithm for finding regular simple paths in outerplanar graphs. *J. Algorithms*, 35(2):235–259, 2000.

[32] F. Olken. Graph data management for molecular biology. *OMICS*, 7(1):75–78, 2003.

[33] C. H. Papadimitriou. *Computational complexity*. Addison-Wesley, 1994.

[34] N. Robertson and P. D. Seymour. Graph Minors .XIII. The Disjoint Paths Problem. *J. Comb. Theory, Ser. B*, 63(1):65–110, 1995.

[35] R. Ronen and O. Shmueli. Soql: A language for querying and creating data in social networks. In *ICDE*, pages 1595–1602, 2009.

[36] A. Schrijver. Finding k Disjoint Paths in a Directed Planar Graph. *SIAM J. Comput.*, 23(4):780–788, 1994.

[37] M. P. Schützenberger. On finite monoids having only trivial subgroups. *Information and Control*, 8(2):190–194, 1965.

[38] L. J. Stockmeyer and A. R. Meyer. Word problems requiring exponential time: Preliminary report. In *STOC*, pages 1–9, 1973.

[39] C. B. Ward and N. M. Wiegand. Complexity results on labeled shortest path problems from wireless routing metrics. *Computer Networks*,

# 7. APPENDIX

**Theorem 3**: *Testing whether a regular language L belongs to trC is:*

1. NL-*complete if L is given by a DFA;*
2. PSPACE-*complete if L is given by an NFA (resp. a regular expression).*

PROOF. 1) Easiness: The proof is based on the characterisation of Lemma 6. Let $A_L = (Q_L, \Sigma, i_L, Q_L, \Delta_L)$ be a DFA that recognizes $L$. We first observe that there is an NL-reduction from this problem to the case where $A_L$ is minimal. The idea is that checking whether two states $q_1$ and $q_2$ are Nerode-equivalent i.e. $\mathscr{L}_{q_1} = \mathscr{L}_{q_2}$ is in NL. For this, we can build on-the-fly the automaton for the language $\mathscr{L}_{q_1} \setminus \mathscr{L}_{q_2} \cup \mathscr{L}_{q_2} \setminus \mathscr{L}_{q_1}$ to check its emptiness.

We assume now that $A_L$ is minimal. We only need to test for each pair of states $q_1$ and $q_2$ whether $q_2 \in \Delta_L(q_1, \Sigma^*)$, $Loop(q_1) \neq \emptyset$ and $Loop(q_2)^M \mathscr{L}_{q_2} \setminus \mathscr{L}_{q_1} = \emptyset$. The first and second statements are easily verified using an NL algorithm for transitive closure. For the third statement, we build a DFA that recognizes $Loop(q_2)^M L(q_2) \setminus L(q_1)$. A DFA for $Loop(q_2)^M \mathscr{L}_{q_2}$ can be built as follow: we make $M$ copies $(A_1, \ldots, A_M)$ of $A_L$. For each copy $A_i$, $i \in [M]$, and each transition from a state $q'$ to $q_2$ inside $A_i$, we replace that transition by a transition from the state $q'$ of $A_i$ to the state $q_2$ of $A_{i+1}$. In the automaton obtained, we choose the state $q_2$ of $A_1$ as the initial state and the final states of $A_M$ as final states. It can be easily checked that the construction can be done in logspace and similarly for the DFA that recognizes $Loop(q_2)^M \mathscr{L}_{q_2} \setminus \mathscr{L}_{q_1}$. As before, the emptiness of this automaton can be checked using an NL algorithm by using transitive closure.

Hardness: we show a reduction from the Emptiness problem. Let $L \subseteq \Sigma^*$ be an instance of Emptiness, given by a DFA $A_L = (Q_L, i_L, Q_L, \Delta_L)$. W.l.o.g, we assume that $\epsilon \notin L$ since this can be checked in constant time. Furthermore, we assume that the symbol 1 does not belong to $\Sigma$. Let $L' = 1^+ L 1^+$. A DFA $A_{L'}$ that recognizes $L'$ can be obtained from $A_L$ as follows. We add a state $q_I$ that will be the initial state of $A_{L'}$ and a state $q_F$ that will be the unique final state of $A_{L'}$. $\Delta_{L'}$ is the extension of $\Delta_L$ defined as follows:

- $\Delta_{L'}(q_I, 1) = q_I$ and $\Delta_{L'}(q_I, a) = i_L$ for every symbol $a \in \Sigma$.
- For every final state $q \in F_L$, $\Delta_{L'}(q, 1) = q_F$.
- $\Delta_{L'}(q_F, 1) = q_F$.

If $L$ is empty then $L' = \emptyset$ belongs to $trC$. Assume that $L$ is not empty. Let $w \in L$. Then, for every $M$, $1^M w 1^M \in L'$ and $1^M 1^M \notin L'$. Thus $L' \notin trC$.

2) Easiness: we first observe the following fact: let $A, B$ be two problems such that $A \in$ NL and let $t$ be a reduction from $B$ to $A$ that works in polynomial space and produces an exponential output. Then $B$ belongs to PSPACE. We can apply this technique here since we can obtain a reduction of that kind from an instance given by an NFA or a regular expression to an equivalent instance given by a DFA. Indeed, this can be achieved using the classical powerset construction for determinization.

Hardness: We use a reduction from the Universality problem. Let $L \subseteq \{0,1\}^*$ be an instance of Universality given by an NFA or a regular expression. Consider $L' = (0+1)^* a^* b a^* + L a^*$ over the alphabet $\{0,1,a,b\}$. Our reduction associates $L'$ to $L$ and keeps the same representation (NFA or regular expression).

If $L = \{0,1\}^*$, then $L' = (0+1)^* a^* (b+\epsilon)$ and thus $L' \in trC$. Conversely, assume $L \neq \{0,1\}^*$. Let $w \in \{0,1\}^* \setminus L$. Then, for every $M$, $wa^M ba^M \in L'$ and $ua^M a^M \notin L'$. Thus $L' \notin trC$. □

**Theorem 4**: *A language $L$ belongs to $trC$ iff $L$ is recognized by a regular expression in $\Psi_{tr}$.*

We prove separately each direction in the next two lemmas.

LEMMA 18. *Every language $L$ in $trC$ can be represented by a regular expression in the above fragment $\Psi_{tr}$.*

PROOF. We next outline an algorithm to build the regular expression $e$ from $A_L$. Let $C_1, \ldots, C_l$ be the strongly connected components of $L$ in some topological order. For every $k \in \{0, \ldots, l\}$ and every sequence $1 \leq j_1 < \cdots < j_k \leq l$, we denote by $L[j_1, \ldots, j_k]$ the set of all words from $L$ that stay for at least $4M^2$ steps in each component $C_{j_1}, \ldots, C_{j_k}$, and stay for at most $4M^2 - 1$ steps in the other components.

Clearly, $L$ is the union of all $L[j_1, \ldots, j_k]$ over all sequences $j_1, \ldots, j_k$. We next show how to build an expression for $L[j_1, \ldots, j_k]$. We denote by $S_1, \ldots, S_k$ the components $C_{j_1}, \ldots, C_{j_k}$ and by $\Sigma_1, \ldots, \Sigma_k$ their alphabet. For any component $S_i$ and state $q$ in $S_i$, we can easily build an expression $H(q)$ for the following language: $H(q) = \{w \in (\Sigma_i)^{2M^2} \mid \exists q_0 \in S_i, \Delta(q_0, w) = q\}$. The rationale for this definition is that when a word $w \in (\Sigma_i)^{M^2} \Sigma_i^*$ is matched from any state of $S_i$, the final state in $S_i$ after matching $w$ is determined by the last $M^2$ letters of $w$, according to Lemma 10: in particular for all $q, q_1 \in S_i$ and $w \in H(q)$, we have $\Delta(q_1, w) = q$.

Let $i < k$ and $q \in S_i$. We build an expression $W(q)$ for the set of all words $w$ that lead from $q$ to some state of $S_{i+1}$ while respecting the sequence of components. In other words, a word $w = a_1 \ldots a_m$ belongs to $W(q)$ iff when we denote by $q_j$ the state $\Delta(q, a_1 \ldots a_j)$, the sequence $q_1 \ldots q_m$ satisfies the following properties: [1]

---
[1] We require somewhat arbitrarily that the first letter of $w$ lets quit $S_i$, while the last letter of $w$ let enter $S_{i+1}$ (i.e., is not in $\Sigma_{i+1}$).

- $q_1 \notin S_i$
- $q_m$ is the first state of the sequence that belongs to $S_{i+1}$
- there are at most $4M^2 - 1$ states $q_j$ in a same component of $A_L$.

$W(q)$ is a finite set of words having length at most $4M^3$.

Similarly, for $i = k$, we build for any state $q \in S_k$ an expression $W(q)$ for the set of all words $w$ that lead from $q$ to some final state while respecting the sequence of components, i.e., satisfying conditions similar to the above ones except that $q_m$ belongs to $F_L$ instead of $S_{i+1}$. $W(q)$ is a finite set of words having length at most $4M^3$.

If $i_L$ belongs to $S_1$, we define the expression $e_{\text{init}}$ as $\epsilon$, otherwise $e_{\text{init}}$ is the set of all words that lead from $i_L$ to some state in $S_1$ while respecting the sequence of components. Rephrased differently, a word $w = a_1 \ldots a_m$ belongs to $W(q)$ iff when we denote by $q_j$ the state $\Delta(i_L, a_1 \ldots a_j)$, the sequence $q_1 \ldots q_m$ satisfies the following properties:

- $q_1 = i_L$
- $q_m$ is the first state of the sequence that belongs to $S_1$,
- there are at most $4M^2 - 1$ states $q_j$ in the same component of $A_L$.

$e_{\text{init}}$ is a finite set of words having length at most $4M^3$.

**Claim 1:** *The expression $e'_0$ defined by the following equations represents the language $L[j_1, \ldots, j_k]$*

$$e'_k = (\Sigma_k)^{\geq 2M^2} \cdot \big(\bigcup_{q \in S_k} H(q) \cdot W(q)\big)$$

$$e'_i = (\Sigma_i)^{\geq 2M^2} \cdot \big(\bigcup_{q \in S_i} H(q) \cdot W(q)\big) \cdot e'_{i+1} \text{ for all } 1 \leq i < k$$

$$e'_0 = e_{\text{init}} \cdot (\Sigma_1)^{\geq 2M^2} \cdot \big(\bigcup_{q \in S_1} H(q) W(q)\big) \cdot e'_1$$

The language of $e'_0$ clearly contains $L[j_1, \ldots, j_k]$. The reciproque follows from the above remark, based on Lemma 10, which concludes the proof of Claim 1. We now define the expressions $e_0, \ldots, e_k$ recursively as follows (with $i$ ranging from 1 to $k$ in the second equation):

$$e_k = ((\Sigma_k)^{\geq 2M^2} + \epsilon) \cdot \big(\bigcup_{q \in S_k} H(q) \cdot W(q)\big)$$

$$e_i = ((\Sigma_i)^{\geq 2M^2} + \epsilon) \cdot \big(\bigcup_{q \in S_i} H(q) \cdot W(q) + \epsilon\big) \cdot e_{i+1}$$

$$e_0 = e_{\text{init}} \cdot ((\Sigma_1)^{\geq 2M^2} + \epsilon) \cdot \big(\bigcup_{q \in S_1} H(q) W(q) + \epsilon\big) \cdot e_1$$

**Claim 2:** *The language of $e_0$ contains $L[j_1, \ldots, j_k]$ and is contained in $L$.*

The language of $e_0$ clearly contains the language of $e_0'$, hence $L[j_1, \ldots, j_k]$ by Claim 1. Let $w \in L(e_0)$. There exist $u_0, v_0, u_1, v_1 \ldots, u_n$, and $v_n$ such that

- $w = u_0 v_0 u_1 v_1 \ldots u_n v_n$
- $u_0 \in L(e_{\text{init}} \cdot ((\Sigma_1)^{\geq 2M^2} + \epsilon))$
- $v_n \in L(\bigcup_{q \in S_k} H(q) \cdot W(q))$
- for each $0 \leq i \leq n-1$, $v_i \in L(\bigcup_{q \in S_i} H(q) \cdot W(q) + \epsilon)$
- for each $1 \leq i \leq n$, $u_i \in L((\Sigma_i)^{\geq 2M^2} + \epsilon)$

Let $w'$ be the word obtained from $w$ by replacing every $v_i$ equal to $\epsilon$ with an arbitrary word from $L(\bigcup_{q \in S_i} H(q) \cdot W(q))$, and every $e_i'$ equal to $\epsilon$ with an arbitrary word from $L((\Sigma_i)^{\geq 2M^2})$. Then $w'$ belongs to $L(e_0')$ and in particular to $L$. Consequently, $w$ also belongs to $L$ by repeated applications of Lemma 11. As $W(q)$ and $H(q)$ are finite sets of words for every state $q$, $e_0$ belongs to the fragment, which concludes the proof of the lemma. □

LEMMA 19. *Let $L$ be a language recognized by a regular expression $\varphi$ in $\Psi_{tr}$. Then $L \in trC$.*

PROOF. Since $trC$ is closed by union, we assume that $\varphi$ is of the form $\varphi_1 \cdot \ldots \cdot \varphi_l$ where $\varphi_1$ and $\varphi_l$ are words and $\varphi_i$ are $\Psi$-terms for every $i \in [1, l-1]$. For each $i \in [l]$, we denote by $L_i$ the language recognized by $\varphi_i$. Let $M$ be the size of $\varphi$ i.e. the number of symbols that compose $\varphi$. Let $u, v, w, w_1, w_2$ be words with $w_1$ and $w_2$ non empty such that $u w_1^M v w_2^M w \in L$. It can be easily seen that there is some term $\varphi_i$ of the form $A^{\geq n} + \epsilon$ such that $w_1 \in A^*$, $u w_1^M \in L_1 \ldots L_i$ and $w_1^M v w_2^M w \in A^{\geq n} \cdot L_{i+1} \ldots L_l$. Similarly, there is some term $\varphi_j$, $j \geq i$ of the form $B^{\geq m} + \epsilon$ such that $w_2 \in B^*$, $u w_1^M v w_2^M \in L_1 \ldots L_j$ and $w_2^M w \in L_j \ldots L_l$. Thus, $u w_1^M w_2^M w \in L_1 \ldots L_{i-1} \cdot A^n A^* \cdot L_j \ldots L_l \subseteq L$. Indeed, if $i = j$ then $A^n A^* L_j \subseteq L_j$ and if $i < j$ then $A^{\geq n} L_j \subseteq L_i L_j$. □

**Theorem 7**: *k-RSPQ is FPT. More precisely, the problem is solvable in time $O(2^{O(k)}|A_L| \cdot |G| \cdot \log |G|)$.*

Let $V$ be a finite set. A $k$-coloring of $V$ is a function $c : V \to [k]$. A set $S \subseteq V$ is colorful for $c$ if $c(x) = c(y) \Rightarrow x = y$ for every $x, y \in S$. The crux of our proof is the following result by Alon et al.:

THEOREM 9 ([2]). *Given $k, n \geq 0$ and a set $V$ of $n$ elements, one can compute in time $O(2^{O(k)}|V|\log|V|)$ a set of $l \in O(2^{O(k)} \log |V|)$ $k$-coloring functions $c_1, \ldots c_l$ such that every set $S$ of $V$ of size $k$ is colorful for at least one $c_i$ $(i \in [l])$.*

PROOF. Let $G, A_L, k$ be an instance of k-RSPQ. We compute $l$ k-coloring functions $c_1, \ldots c_l$ as stated in Theorem 9. Let $c$ one of these functions. We will show how to decide if there is a colorful $L$-labeled path from $x$ to $y$ in $(G, c)$. To this purpose, we define a function $f : V \times Q_L \times \mathcal{P}([k]) \to \{0, 1\}$ such that $f(v, q, S) = 1$ if there exists a colorful path $p$ starting from $x$ that uses only colors of $S$ and such that $\Delta_L(i_Q, w) = q$ where $w$ is the label of $p$. Clearly there is a colorful path from $x$ to $y$ if there is a set $S \subseteq [k]$ and a final state $q \in F_L$ such that $f(y, q, S) = 1$.

The function can be computed by dynamical programming using the following equation.

- $f(x, i_Q, \{c(x)\}) = 1$
- $f(v, q, S) = 1$ if there is a subset $S' \subsetneq S$ such that $f(v, q, S') = 1$;
- $f(v, q, S) = 1$ if $c(v) \in S$ and there is an edge $v'$, a state $q'$ and a label $a$ such that $f(v', q', S \backslash c(v)) = 1$, $(v, a, v') \in E$ and $q' \in \Delta_L(q, a)$;
- $f(v, q, S) = 0$ otherwise.

This function can be computed in time $O(2^k \cdot |A_L| \cdot |G|)$. We compute $f$ for every function $c_i$, $i \in [l]$ where $l \in O(2^{O(k)} \log |V|)$. Consequently, k-RSPQ can be solved in time $O(2^{O(k)}|A_L| \cdot |G| \cdot \log |G|)$. □

**Theorem 8**: *Let $k \geq 0$ and $\mathcal{G}$ be a class of db-graphs of directed treewidth at most $K$. Then, $RSPQ(Reg, \mathcal{G})$ is polynomial, where Reg denotes the regular languages.*

PROOF SKETCH. The proof is a straighforward adaptation of the proof proposed in [23] for the Hamiltonian Path problem. Since they use a dynamic approach, they consider a more general problem: given a digraph $G$ and a sequence of k tuples $(v_i, n_i, v_i')_{i \in [k]}$, are there $k$ disjoint simple paths $p_1, \ldots p_k$ such that $p_i$ is a path of size $n_i$ from $v_i$ to $v_i'$ for every $i \in [k]$?

We extend the problem as follows: given a db-graph $G$, a regular language $L$ and a sequence of k tuples $(v_i, n_i, v_i', q_i, q_i')_{i \in [k]}$, are there $k$ words $w_1, \ldots w_k$ and $k$ disjoint simple paths $p_1, \ldots p_k$ such that $p_i$ is a $w_i$-labeled path of size $n_i$ from $v_i$ to $v_i'$ and $\Delta_L(q_i, w_i) = q_i'$ for every $i \in [k]$? Therefore, their proof can easily be adapted to this new problem. □